\documentclass[aps,pra,twocolumn,superscriptaddress,raggedbottom,showpacs,floatfix,longbibliography]{revtex4-1}
\usepackage{amsmath,amsfonts,amssymb,amsthm}
\usepackage{bm}
\usepackage{color}
\usepackage{graphicx,latexsym}
\usepackage{epstopdf}
\usepackage{appendix}

\begin{document}
\renewcommand{\vec}{\mathbf}
\renewcommand{\Re}{\mathop{\mathrm{Re}}\nolimits}
\renewcommand{\Im}{\mathop{\mathrm{Im}}\nolimits}

\title{Real space mean-field theory of a spin-1 Bose gas in synthetic dimensions}
\author{Hilary M. Hurst}
\affiliation{Condensed Matter Theory Center and Joint Quantum Institute, Department of Physics, University of Maryland, College Park, Maryland 20742, USA}
\author{Justin H. Wilson}
\affiliation{Institute of Quantum Information and Matter and Department of Physics, California Institute of Technology, Pasadena, CA 91125 USA}
\affiliation{Condensed Matter Theory Center and Joint Quantum Institute, Department of Physics, University of Maryland, College Park, Maryland 20742, USA}
\author{J. H. Pixley}
\affiliation{Condensed Matter Theory Center and Joint Quantum Institute, Department of Physics, University of Maryland, College Park, Maryland 20742, USA}
\date{\today}
\author{I. B. Spielman}
\affiliation{Joint Quantum Institute, National Institute of Standards and Technology,
and University of Maryland, Gaithersburg, Maryland, 20899, USA}
\author{Stefan S. Natu}
\affiliation{Condensed Matter Theory Center and Joint Quantum Institute, Department of Physics, University of Maryland, College Park, Maryland 20742, USA}

\begin{abstract}
The internal degrees of freedom provided by ultracold atoms give a route for realizing higher dimensional physics in systems with limited spatial dimensions. Non-spatial degrees of freedom in these systems are dubbed ``synthetic dimensions".  This connection is useful from an experimental standpoint but complicated by the fact that interactions alter the condensate ground state. Here we use the Gross-Pitaevskii equation to study ground state properties of a spin-1 Bose gas under the combined influence of an optical lattice, spatially varying spin-orbit coupling, and interactions at the mean-field level. The associated phases depend on the sign of the spin-dependent interaction parameter and the strength of the spin-orbit field. We find ``charge" and spin density wave phases which are directly related to helical spin order in real space and affect the behavior of edge currents in the synthetic dimension. We determine the resulting phase diagram as a function of the spin-orbit coupling and spin-dependent interaction strength, considering both attractive (ferromagnetic) and repulsive (polar) spin-dependent interactions, and we provide direct comparison of our results with the non-interacting case. Our findings are applicable to current and future experiments, specifically with $^{87}$Rb, $^{7}$Li, $^{41}$K, and $^{23}$Na.
\end{abstract}

\maketitle

\section{Introduction}
Internal degrees of freedom in atomic Bose-Einstein condensates (BECs) provide a platform for realizing phenomena conceived of in more traditional condensed matter settings. We view these discrete internal spin degrees of freedom as an extra ``synthetic" dimension with finite extent, allowing phenomena in higher dimensions to exist in systems  with lower real space dimension~\cite{Celi2014}. The setup considered here consists of a one-dimensional (1D) spin-1 Bose gas in an optical lattice potential where the three hyperfine levels are ``Raman" coupled using a pair of laser beams, a scheme which has been explored both theoretically and experimentally~\cite{Celi2014,Luo2015,Kolley2015,Atala2014,Zeng2015,Stuhl2015,Mancini2015}.

\begin{figure}[t!]
\centering
\includegraphics[width =\columnwidth]{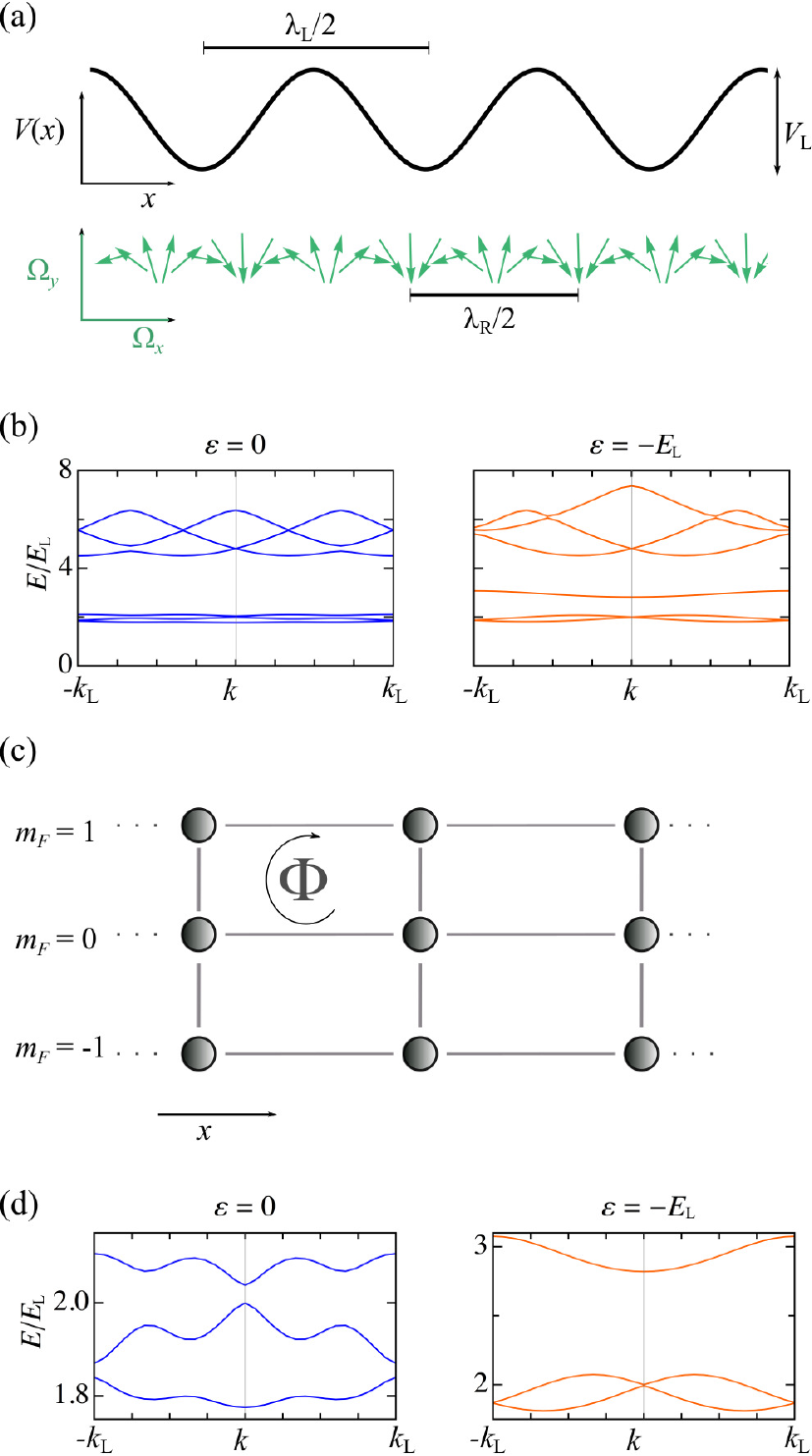}
\caption{(color online)~(a)~The physical system consists of an optical lattice (black line, period $\lambda_\mathrm{L}/2$) with Raman lasers forming an effective helical magnetic field (green arrows, period $\lambda_\mathrm{R}/2$).~(b)~Single particle dispersion relation for the spin-1 spin-orbit coupled Bose gas in an optical lattice at $\Omega = 0.25E_\mathrm{L}$, $c_2 = 0$, $c_0 = 0$. The six lowest energy bands are pictured, with the three lowest bands well split from the higher energy modes.~(c)~Synthetic dimensions visualization. They hyperfine levels $m_F$ are viewed as an additional dimension with $2F+1$ sites. Each plaquette has a uniform flux $\Phi \approx 2\pi k_\mathrm{R}/k_\mathrm{L}$.~(d)~Three lowest bands in the synthetic dimensions set up for $\Omega = 0.25E_\mathrm{L}$. At small $\Omega$, $\varepsilon = 0$ the bottom band has three minima, with the lowest energy minimum at $k=0$. For $\varepsilon < 0$ the bottom band has two degenerate minima, reflecting degeneracy in $m_F  =\pm 1$.\label{Fig:Bands}}
\end{figure}

\begin{figure}[t!]
\centering
\includegraphics[width = \columnwidth]{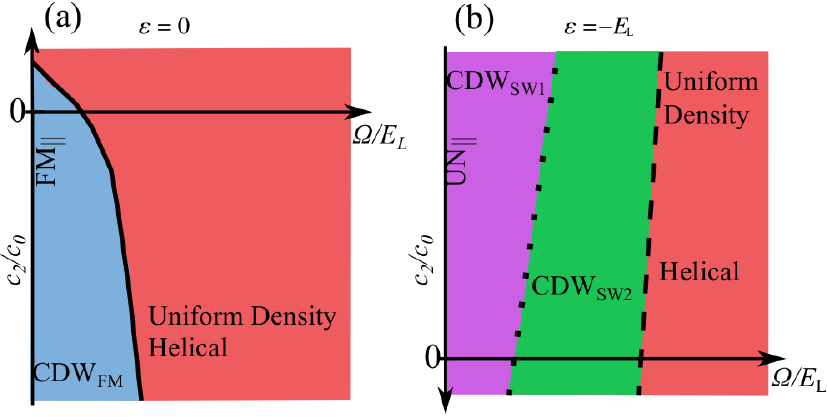}
\caption{~(color online)~Schematic phase diagram for $\varepsilon = 0$ and $\varepsilon = -E_\mathrm{L}$.~(a)~$\varepsilon = 0$. For $\Omega \lesssim 0.5E_\mathrm{L}$ the system exhibits charge density wave behavior and spin polarization along $S_x$ and is denoted CDW$_\mathrm{FM}$. Increasing $\Omega$ leads to a uniform density phase with a helical spin texture. The period of the spin helix is determined by the Raman field. Positive $c_2$ values suppress density fluctuations.~(b)~$\varepsilon = -E_\mathrm{L}$. The BEC exhibits distinct charge density wave phases with different ordering wavevectors and different spin textures, denoted CDW$_\mathrm{SW1}$ and CDW$_\mathrm{SW2}$. A cross over occurs between the two with increasing $\Omega$. At $\Omega \approx 2.4E_\mathrm{L}$ there is a first order transition to a uniform density state with helical spin polarization.\label{Fig:Main}}
\end{figure}

Experimental advances in ultra-cold atomic gases led to spin-orbit coupling (SOC) in spinful Bose and Fermi gases, one route for realizing synthetic dimensions~\cite{Galitski2013,Beeler2013,Dalibard2011,Stamper2013}. Despite the lack of true Bose-Einstein condensation in quasi-1D, in the weakly interacting mean-field (MF) regime the condensate wave function is well described by the 1D Gross-Pitaevskii equation (GPE)~\cite{Stamper2013}. The introduction of a spin-orbit wave vector imbues the single particle energy dispersion with multiple minima in momentum space~\cite{Wang2010,Ho2011,Lan2014}. At low temperatures an interacting Bose gas can Bose-condense at these minima, forming a superfluid (SF) with density order: a charge density wave (CDW)~\cite{Wang2010,Ho2011,Cong2011,Li2012}. Moreover, different SF phases occur depending on the symmetry of the underlying Hamiltonian: spin density wave (SW) and magnetized phases are two examples~\cite{Xu2012}. In spin-$1/2$ bosons, SOC can induce CDW and SW phases, however these are necessarily pseudospin systems and an SU(2)-breaking spin dependent interaction term is required to achieve these phases~\cite{Li2012,Yu2013}. For the case of spin-1 bosons, spin-dependent interactions preserve SU(2) symmetry which is then broken by SOC, leading to a rich phase diagram exhibiting multiple CDW and SW phases~\cite{Lan2014,Natu2015,Sun2015,Martone2015}.

The second ingredient to the synthetic dimension programme is an optical lattice. The system is loaded into a 1D lattice provided by counter propagating lasers with wavelength $\lambda_\mathrm{L} = 2\pi/k_\mathrm{L}$ where $k_\mathrm{L}$ is the recoil momentum. The hyperfine spin states $-F\leq m_F\leq F$ are viewed as an added spatial dimension, coupled using Raman lasers with a different wavelength $\lambda_\mathrm{R} = 2\pi/k_\mathrm{R}$. These components are shown in Fig.~\ref{Fig:Bands}(a). Thus, the 1D system maps to a two dimensional ladder model with rungs of $2F+1$ sites in width, leading to a square lattice in the tight binding approximation [see Fig.~\ref{Fig:Bands}(c)]. The spatial dependence of the Raman coupling is essential to this analogy, as it gives each synthetic plaquette a flux $\Phi = 2\pi k_\mathrm{R}/k_\mathrm{L}$~\cite{Celi2014}. In this space, the laser coupling of spin states gives hopping along the synthetic dimension direction. This allows for novel transport properties and topological states of matter to form and be probed~\cite{Celi2014,Atala2014,Price2015,Stuhl2015,Mancini2015}. This system was theoretically investigated for several different types of atoms and recent experiments observed chiral currents~\cite{Celi2014,Luo2015,Kolley2015,Zeng2015,Atala2014,Stuhl2015}. The mapping to a higher dimensional Hamiltonian is exact for single particle physics, but local interactions in the 1D system translate to non-local interactions in the synthetic direction. In this work, we explore the combined effect of the Raman strength and spin-dependent interactions on phases at the MF level, without making tight binding or single band approximations. In particular, we focus on the regime of intermediate lattice depth where the mean-field description is applicable. At higher lattice depths, Mott physics becomes important and the GPE is an insufficient probe of the system. All of the parameter values used in our calculations are listed in Table \ref{tab:notation}.

\begin{figure}[t!]
\centering
\includegraphics[width =\columnwidth]{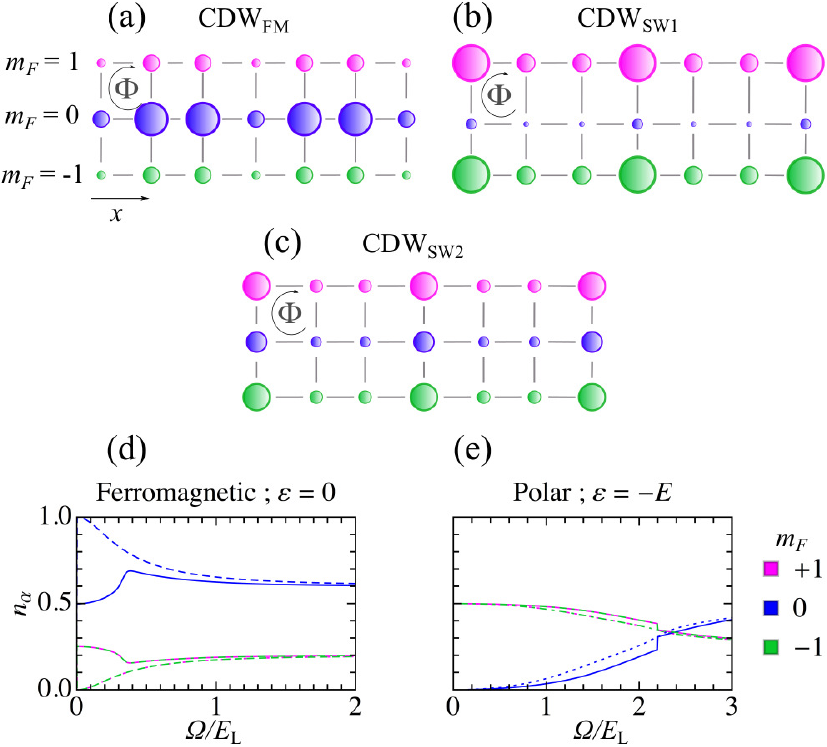}
\caption{(color online)~(a)~Schematic diagram of the CDW$_{\mathrm{FM}}$ phase. The BEC predominantly occupies $m_F=0$ level. The total density is modulated at neighboring sites due to the Raman field.~(b)~Schematic diagram of the CDW$_\mathrm{SW1}$ phase. The edges are preferentially occupied and there is an overall density modulation.~(c)~Schematic diagram of the CDW$_\mathrm{SW2}$ phase. The bulk is more occupied than in (b), and the overall density modulation remains.~(d-e)~Fractional population as a function of $\Omega$.~(d)~$c_2/c_0=-0.25$, $\varepsilon=0$; The system begins in a CDW$_\mathrm{FM}$ ground state at $\Omega\approx0$ with $n_0 = 1/2$ and $n_{\pm 1} = 1/4$ and moves to meet the single particle occupation, indicated by dashed lines.~(e)~$c_2/c_0 = 0.25, \varepsilon = -E_\mathrm{L}$. The system undergoes an edge to bulk first order transition at $\Omega \approx 2.0E_\mathrm{L}$, which is weakened to a cross over in the limit $c_2=0$. As $\Omega$ increases the bulk is preferentially occupied. Dotted lines indicate the case for $c_0 \neq 0, c_2 = 0$. 
\label{Fig:Fractions}}
\end{figure}

Previous work identified numerous MF phases without the optical lattice~\cite{Natu2015,Yu2016,Chen2016,Martone2015}, and with a deep optical lattice resulting in pinning effects and an interaction driven SF phase~\cite{Pixley2015}. ``Pinning" refers to condensation only at wavevectors commensurate with the underlying lattice~\cite{Pixley2015}. The effect of increasing the lattice depth (that is intermediate between these two regimes) was also recently explored for spin-1/2 systems~\cite{Chen2016b,Martone2016}. A common feature of these systems is that the lattice causes the condensation at the Brillouin zone edge, which coincides with the wavevector of the optical lattice potential and not with the wavevector of spin-orbit coupling~\cite{Pixley2015,Chen2016b,Martone2016}.

The physics that continuously connects the continuum limit to the deep lattice limit (i.e. both the single particle Hamiltonian, optical lattice, and tight binding) is largely unexplored for spin-1 spin-orbit coupled bosons. In light of recent experimental progress it is important to understand the possible ground state phases of this system including interactions to compare with all parameter ranges possible in experiment. We study the ground state properties of the spin-1 Bose gas with SOC and an optical lattice at the mean-field level by solving the GPE at zero temperature for weakly interacting bosons with either repulsive (polar) or attractive (ferromagnetic) spin-dependent interactions and repulsive density dependent interactions. Of particular interest is how the phases develop with increasing Raman coupling and how different phases manifest in the synthetic dimensions picture. Furthermore, we compare how the synthetic dimensions set up affects previously studied phenomena in the uniform system, such as the appearance of CDW phases~\cite{Wang2010,Lan2014,Natu2015}.

\begin{table*}[t!]
    \centering
    \begin{tabular}{c|c|c}
        \hline \hline
        Notation & Description & Value \\
        \hline
         $\hat{\psi}_\alpha(x)$ & Boson field operator; $\alpha = m_F$ & -- \\
         $N$ & Number; $N = \int dx \, \langle{\hat\psi^\dagger_\alpha \hat\psi_\alpha}\rangle$. & $N=100$ \\
         $L$ & length of lattice & $L = 15\lambda_\mathrm{L}/2$  \\
         $M$ & atomic mass  & -- \\
         $V(x)$ & Optical lattice potential; $V(x) = V_\mathrm{L} \cos^2(k_\mathrm{L} x)$ & $V_\mathrm{L} = 5 E_\mathrm{L}$\\
         $k_\mathrm{L}$ & Lattice recoil momentum & $2\pi/\lambda_\mathrm{L}$ \\
         $E_\mathrm{L}$ & Lattice recoil energy $E_\mathrm{L} = k_\mathrm{L}^2/2M$ & -- \\
         $v_\mathrm{L}$ & Lattice recoil velocity $v_\mathrm{L} = k_\mathrm{L}/M$ & -- \\
         $\varepsilon$ & Quadratic Zeeman strength & $\varepsilon = \begin{cases}0, & c_2<0, \\ -E_\mathrm{L}, & c_2>0. \end{cases} $ \\
         $\mathbf{F}_{\alpha\beta}$ & vector of spin-1 matrices & -- \\
         $\bm{\Omega}_\mathrm{R}(x)$ & spin-orbit coupling; $\bm{\Omega}_\mathrm{R}(x) = \Omega \Re\left[e^{2i k_\mathrm{R} x}(\hat{\mathbf{x}} + i \hat{\mathbf{y}})\right]$ & $\Omega$ tuned \\
         $k_\mathrm{R}$ & Raman wave vector & $k_\mathrm{R}/k_\mathrm{L}=4/3$. \\
         $c_0$ & density-density interaction & $c_0 N/L = 0.1 E_\mathrm{L}$ \\
         $c_2$  & spin-dependent interaction & $\frac{c_2}{c_0} = \begin{cases} \scriptstyle{-0.25,-0.5,-0.7}\\ \scriptstyle{0.25,0.5,1.0} \end{cases}$ \\
         \hline\hline
    \end{tabular}
    \caption{Table of notation and values used for numerical simulations (if applicable). We tune the Raman field strength $\Omega$ and spin-dependent interaction $c_2$. Values for $V_\mathrm{L}$ and $k_\mathrm{R}$ come from the relevant experiment~\cite{Stuhl2015}. Interactions are related to scattering lengths $a_0$ and $a_2$ by $c_0 = 4\pi(a_0+2a_2)/3M$ and $c_2 = 4\pi(a_2 - a_0)/3M$~\cite{Stamper2013}.}
    \label{tab:notation}
\end{table*}

The main result of this work is shown in Fig.~\ref{Fig:Main}, the phase diagrams for varying Raman strength $\Omega$ and ferromagnetic and polar spin-dependent interaction parameter $c_2$ for two different values of the quadratic Zeeman strength. We briefly discuss our conclusions here, with more detail provided in sections III and IV. As expected from previous work~\cite{Pixley2015}, the lattice suppresses condensation at wavevectors other than $k=0$ and the lattice wavevector $k = k_\mathrm{L}$ at the Brillouin zone edge. However, in the regime of interest this order is not completely suppressed and we predict several novel phases. Phases are labeled CDW or uniform density, with the type of spin texture denoted by the subscript in the CDW regimes. When the interaction strength is comparable to the Raman coupling, along with the CDW phases we find a variety of spin textures: a predominantly ferromagnetic state (FM) for $c_2 <0$ and two different spin density waves (SW1, SW2) for $c_2 >0$. Increasing $\Omega$ favors the single-particle ground state with uniform density and helical spin order. For the purposes of this work, a ``uniform density state" refers to a state where the total density is modulated only by the optical lattice. We relate these phases to the synthetic dimensions picture by analyzing the spin current and fractional population of atoms in each spin state. The rich variety of phases reported here directly results from the interplay between interactions and the single particle Hamiltonian. We are able to establish what effects result directly from interactions by comparing our results with exact results for the ground state in the non-interacting case. 

Our results align with MF phases previously studied without the lattice, and we conclude that the intermediate lattice depth modifies the phase boundaries but does not destroy phases that have already been predicted~\cite{Chen2016,Yu2016,Natu2015}. In addition, we characterize the lattice depth at which these phases are suppressed and the mean-field picture breaks down. Furthermore, we find that the ground state phase is dependent on the strength of the Raman coupling, which provides an additional tunable parameter in experiment. It is notable that increasing the Raman coupling strength at constant lattice depth leads to condensation at the Brillouin zone edge, a phenomenon that was previously predicted solely for increasing lattice depth~\cite{Martone2016}. Our model parameters were selected to be directly relevant to experiments with ferromagnetic atoms (e.g. $^{87}$Rb, $^{7}$Li, $^{41}$K) as well as polar atoms (e.g. $^{23}$Na)~\cite{Stuhl2015,Mancini2015,Zeng2015}. In Section II we explain the model and briefly review previous results. Sections III and IV give detailed results for attractive and repulsive spin-dependent interactions, respectively. We conclude in Section V and discuss how our work relates to current and future experiments. The effects of increasing lattice depth are provided in the Appendix.

\section{Model and Method}

We consider interacting spin-1 bosons in the presence of a Raman field and an optical lattice. The model is defined by
$\hat{H} = \int dx(\hat{\mathcal{H}}_0 + \hat{\mathcal{H}}_\mathrm{so} + \hat{\mathcal{H}}_\mathrm{int})$, setting $\hbar = 1$ throughout. The non-interacting Hamiltonian density $\hat{\mathcal{H}}_0$ is 
\begin{equation}
\hat{\mathcal{H}}_0 = \frac{1}{2M}\nabla\hat{\psi}_\alpha^\dagger\nabla\hat{\psi}_\alpha+ \hat{\psi}_\alpha^\dagger\left[ V(x)- \mu \right] \hat{\psi}_{\alpha}, 
\label{eqn:H0}
\end{equation}
with the hyperfine states coupled through the spin-orbit coupling and interaction terms 
\begin{align}
\hat{\mathcal{H}}_{\mathrm{so}}&=\hat{\psi}^\dagger_\alpha \left[ \varepsilon(F^2_z)_{\alpha\beta} + \boldsymbol{\Omega}_\mathrm{R}(x) \cdot \vec{F}_{\alpha\beta}\right] \hat{\psi}_\beta,\label{eqn:H1}\\
\hat{\mathcal{H}}_{\mathrm{int}}&=  
\frac{c_0}{2} \hat{\psi}^\dagger_\alpha \hat{\psi}^\dagger_\beta  \hat{\psi}_\beta  \hat{\psi}_\alpha 
+ \frac{c_2}{2} \hat{\psi}^\dagger_\gamma \hat{\psi}^\dagger_\alpha   \vec{F}_{\gamma \nu}\cdot\vec{F}_{\alpha\beta}
\hat{\psi}_\beta  \hat{\psi}_\nu,
\label{eqn:H2}
\end{align}
using the notation defined in Table~\ref{tab:notation}, with repeated indices summed over. In particular, the spatial structure of the the spin-orbit coupling term is given by $\boldsymbol{\Omega_\mathrm{R}}(x) = \Omega\cos(2k_\mathrm{R}x)\vec{e}_x -\Omega\sin(2k_\mathrm{R}x)\vec{e}_y$.

The single particle physics of this system without a lattice was studied extensively in Ref.~\cite{Lan2014}. For small $\Omega$ and $\varepsilon$ the low energy dispersion relation has three minima at $k=0, \pm~2 k_\mathrm{R}$ corresponding to spin states $m_F=0,\pm1$. These minima are degenerate when $\varepsilon$ is tuned to slightly negative values with increasing $\Omega$. Increasing $\varepsilon$ shifts the middle minimum down, resulting in a single minimum structure, while decreasing $\varepsilon$ shifts the middle minimum up, leading to a double minima structure. The dispersion relation for these two conditions is shown in Fig.~\ref{Fig:Bands}(b,d), where we plot the band structure for $\Omega = 0.25E_\mathrm{L}$ and $\varepsilon = 0$ and $-E_\mathrm{L}$. In the triply-degenerate regime, the condensate wavefunction takes the form~\cite{Natu2015}
\begin{equation}
    \psi(x) = A_+\xi_+e^{i2k_\mathrm{R}x}+A_0\xi_0+A_-\xi_-e^{-i2k_\mathrm{R}x}, 
    \label{eqn:threeminima}
\end{equation}
where $A_{\pm,0}$ are complex amplitudes and $\xi_{\pm,0}$ are the single-particle spinor eigenstates at the energy minima corresponding to momenta $k = \pm 2k_\mathrm{R}, 0$. The condensate can exhibit a zero momentum phase or a plane wave phase when a single minimum is occupied, corresponding to $A_0 \neq 0, A_\pm = 0 $ or $A_\pm \neq 0, A_{\mp,0} = 0$. The condensate also exhibits various density modulated (CDW) phases when at least two of the three components $A_{\pm,0}$ are nonzero. These CDW phases have different wavelengths depending how the minima are occupied~\cite{Lan2014}.

\begin{figure}[t!]
\includegraphics[width=\columnwidth]{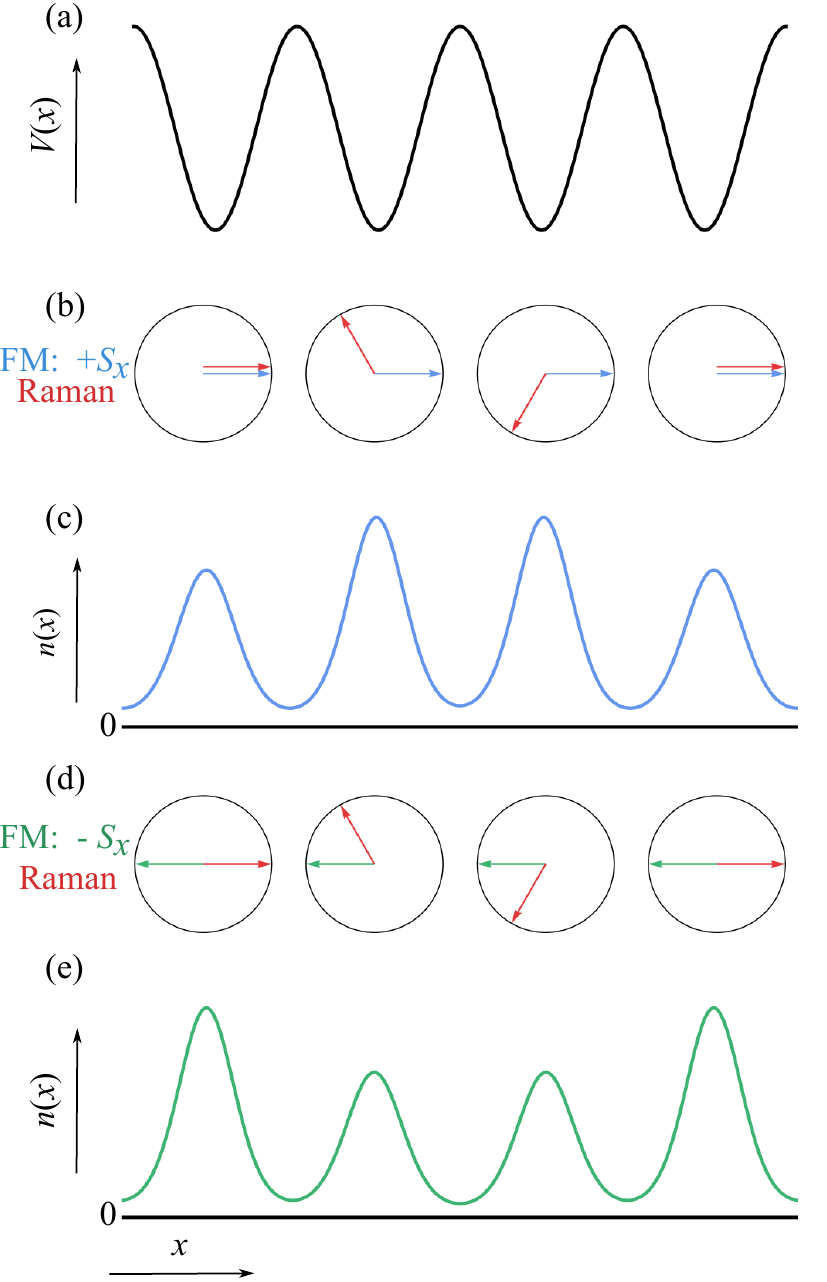}
\caption{(color online)~Interplay of spin and density for $c_2 < 0$ at small $\Omega$.~(a)~Lattice potential $V(x)$.~(b) Local spin polarization and Raman field at each site for $+S_x$ polarization. The blue arrow shows the spin polarization, and the red arrow shows the local Raman field.~(c)~CDW$_\mathrm{FM}$ phase with $+S_x$ polarization. Density increases at sites where $\vec{F}\cdot\boldsymbol{\Omega}_\mathrm{R} < 0$.~(d)~Local spin polarization and Raman field at each site for $-S_x$ polarization. The green arrow shows the spin polarization, and the red arrow shows the local Raman field.~(e)~CDW$_\mathrm{FM}$ phase with $-S_x$ polarization. Density increases at the sites where $\vec{F}\cdot\boldsymbol{\Omega}_\mathrm{R} < 0$.\label{Fig:CDWSketch}}
\end{figure}

For a weakly interacting condensate Eq.~\eqref{eqn:threeminima} is still a valid ansatz, but $A_{\pm,0}$ are selectively occupied to minimize both the single particle and interaction energies. Interactions dictate the form of the spinor structure in the condensate, favoring ferromagnetic order for attractive ($c_2<0$) and uniaxial nematic order for repulsive ($c_2>0$) spin-dependent interactions~\cite{Natu2015}. This is due to the fact that for $c_2 < 0~(>0)$, the system maximizes (minimizes) spin $\langle\hat{\vec{S}}(x)\rangle$, where $\hat{\vec{S}}(x) = \hat{\psi}^\dagger_\alpha(x)\vec{F}_{\alpha\beta}\hat{\psi}_\beta(x)$, leading to distinct phases in the two regimes~\cite{Lan2014,Natu2015,Martone2015,Yu2016}. Furthermore, tuning $\varepsilon$ also alters the ground state in the presence of interactions by changing the structure of the underlying dispersion relation~\cite{Natu2015}. This interplay between SW and CDW order leads to a number of exotic phases and excitations in the continuum system~\cite{Natu2015,Sun2015,Martone2015,Yu2016}. 

One goal of the present manuscript is to understand the stability of each spin-orbit and interaction driven phase in the presence of an optical lattice, away from the deep-lattice limit. A lattice invalidates the ansatz of Eq.~\eqref{eqn:threeminima} since the lattice breaks translational symmetry, but essential features and minima of the lowest band remain intact as shown in Fig.~\ref{Fig:Bands}(b,d).

\begin{figure}[t!]
\centering
\includegraphics[width =\columnwidth]{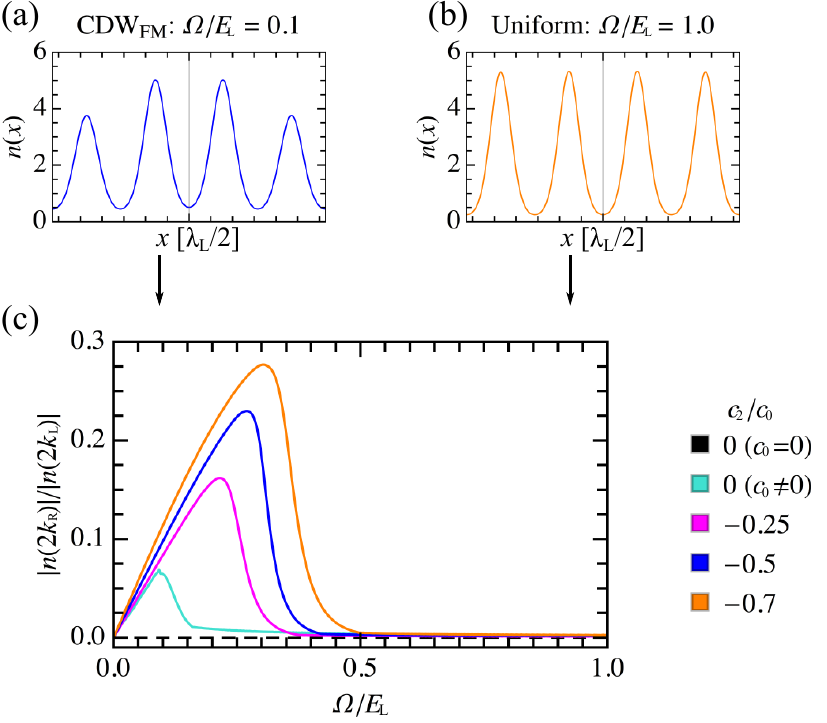}
\caption{(color online)~(a-b)~GPE density computed for $c_2/c_0 = -0.25$ and $\varepsilon = 0$.~(a)~Density in real space in the CDW$_\mathrm{FM}$ phase shows modulation between lattice sites.~(b)~Density in real space in the uniform density phase.~(c)~$|n(2k_\mathrm{R})| \neq 0$ signals a CDW phase. The non-interacting case (dashed line) has density modulation only from the lattice, and only slight density modulation appears for $c_2 = 0, c_0 > 0$. Increasing spin-dependent interaction strength $|c_2|$ leads to greater overall density modulation until a cross over occurs to the uniform density regime.\label{Fig:DensityF}}
\end{figure}

We describe a spinor BEC by three classical complex fields $\langle \hat{\psi}_\alpha(x) \rangle =\psi_\alpha(x) = \sqrt{n(x)}\xi_\alpha$ where $n(x) = \sum_\alpha |\psi_\alpha|^2$ is the total density, $\xi_\alpha$ is a three component spinor with normalization $\xi^*_\alpha \xi_\alpha = 1$, and $\alpha = \left\{1,0,-1\right\}$ labels the synthetic dimension sites. We define the GPE energy functional by replacing the bosonic operators in Eqs.~\eqref{eqn:H0}-\eqref{eqn:H2} with $\psi_\alpha(x)$. This gives the coupled equations
\begin{multline}
i\partial_t\psi_\alpha = \left\{\left[-\frac{\nabla^2}{2M} + V(x)\right]\delta_{\alpha\beta}+\boldsymbol{\Omega_\mathrm{R}}\cdot\vec{F}_{\alpha\beta}\right. \\
+ \varepsilon(F_z^2)_{\alpha\beta} \left.+ c_0|\psi|^2\delta_{\alpha\beta}+c_2\left[(\psi^\dagger_{\delta}\vec{F}_{\delta\gamma}\psi_\gamma)\cdot\vec{F}_{\alpha\beta}\right]\right\}\psi_{\beta}
\label{eqn:EOM}
\end{multline}

We solve for the ground state using imaginary-time evolution where $t\rightarrow -i\tau$~\cite{Chiofalo2000} and test convergence using the strong criterion detailed in Ref.~\cite{Antoine2014}. The system was initialized in a uniform state $\sum_\alpha|\psi_\alpha(x)|^2 = \mu/c_0$, with all 3 spin components (ladder legs) equally weighted and where the number of particles fixes $\mu$. We find that near the phase transitions there are several states that are close in energy. Our initial state biases the GPE solver and at some point in the phase diagram there is an artificial transition between phases, because sometimes the GPE relaxes to a metastable state rather than the ground state. To precisely pinpoint a transition, we ran the GPE over parameter ranges biasing the initial guess in favor of the previously calculated result at $\Omega\pm \delta \Omega$ (running from small to large $\Omega$ and large to small $\Omega$). The lower energy phase is then taken as the ground state. As a heuristic check, we stochastically sampled points in the phase diagram using random initial conditions, confirming the ground state solution found from the above methodology. The parameter values used in the GPE solver are detailed in Table~\ref{tab:notation}. Importantly, $\varepsilon$ has a strong influence on phases and we present results for ferromagnetic BECs with $\varepsilon = 0$ and polar BECs with $\varepsilon = -E_\mathrm{L}$. These two choices are explained in Sections III and IV. 

CDW and SW phases are identified by nonzero Fourier amplitudes of the density and spin order parameters, $n(k)$ and $\langle \vec{S}(k)\rangle$, at the relevant wavevectors $k=2k_\mathrm{L}$, $k=2k_\mathrm{R}$, and $k=4k_\mathrm{R}$. The first two correspond to the wavevector of the lattice and the spin-orbit field, respectively. Nonzero amplitude at $k = 2k_\mathrm{L}$ indicates an effect due to the lattice, while $k = 2k_\mathrm{R}$ indicates an effect due to SOC. The third wavevector, $k=4k_\mathrm{R}$, corresponds to condensation at the two degenerate minima at $\pm 2k_\mathrm{R}$ in the single particle bandstructure (see Fig.~\ref{Fig:Bands}). The resolution of the system is set by the length $L$; here we have 5 unit cells because the Raman beam is periodic over 3 optical lattice sites. Density $n(k)$ and spin $\langle\vec{S}(k)\rangle$ are defined by $n(k) = \int dx~e^{ikx}n(x)$ and $\langle\vec{S}(k)\rangle = \int dx~e^{ikx}\langle\vec{S}(x)\rangle$. A schematic diagram of CDW phases in synthetic dimensions is shown in Fig.~\ref{Fig:Fractions}(a-c). In synthetic space, these phases are captured by the fractional population in the $m_F$ states, shown in Figs.~\ref{Fig:Fractions}(d-e). Fractional population $n_{\alpha}$  is defined as 
\begin{equation}
    n_{\alpha} = \frac{\int dk \, n_{\alpha}(k)}{\sum_{\alpha} \int dk \, n_{\alpha}(k)}.
\end{equation}

Finally, we analyze the spin currents in this system, which is analogous to the chiral edge current in a quantum Hall system. The extremal spins represent the edges in the synthetic dimension~\cite{Beeler2013}. This provides a way to visualize and measure chiral currents because the BEC can be imaged in the synthetic and spatial dimensions using spin-resolved absorption imaging~\cite{Stuhl2015}. The total spin current density is defined as 
\begin{align}
    j_\mathrm{S} \equiv \sum_{\alpha} \alpha \langle j_{\alpha} \rangle = \sum_{\alpha} \alpha \int \frac{d k}{2\pi} \frac{k}{M} \, \psi^\dagger_{\alpha}(k)\psi_{\alpha}(k),
\end{align}
where $\psi_\alpha(k) = \int dx~e^{ikx}\psi_\alpha(x)$. Nonzero current corresponds to occupying states in the edge-site conduction bands of the corresponding 2D lattice system. In this case, the spin current is driven by the Raman beam $\Omega$, and also depends on the population of atoms in the $m_F = \pm1$ Zeeman (edge) states.

\begin{figure}[t!]
\centering
\includegraphics[width = 0.9\columnwidth]{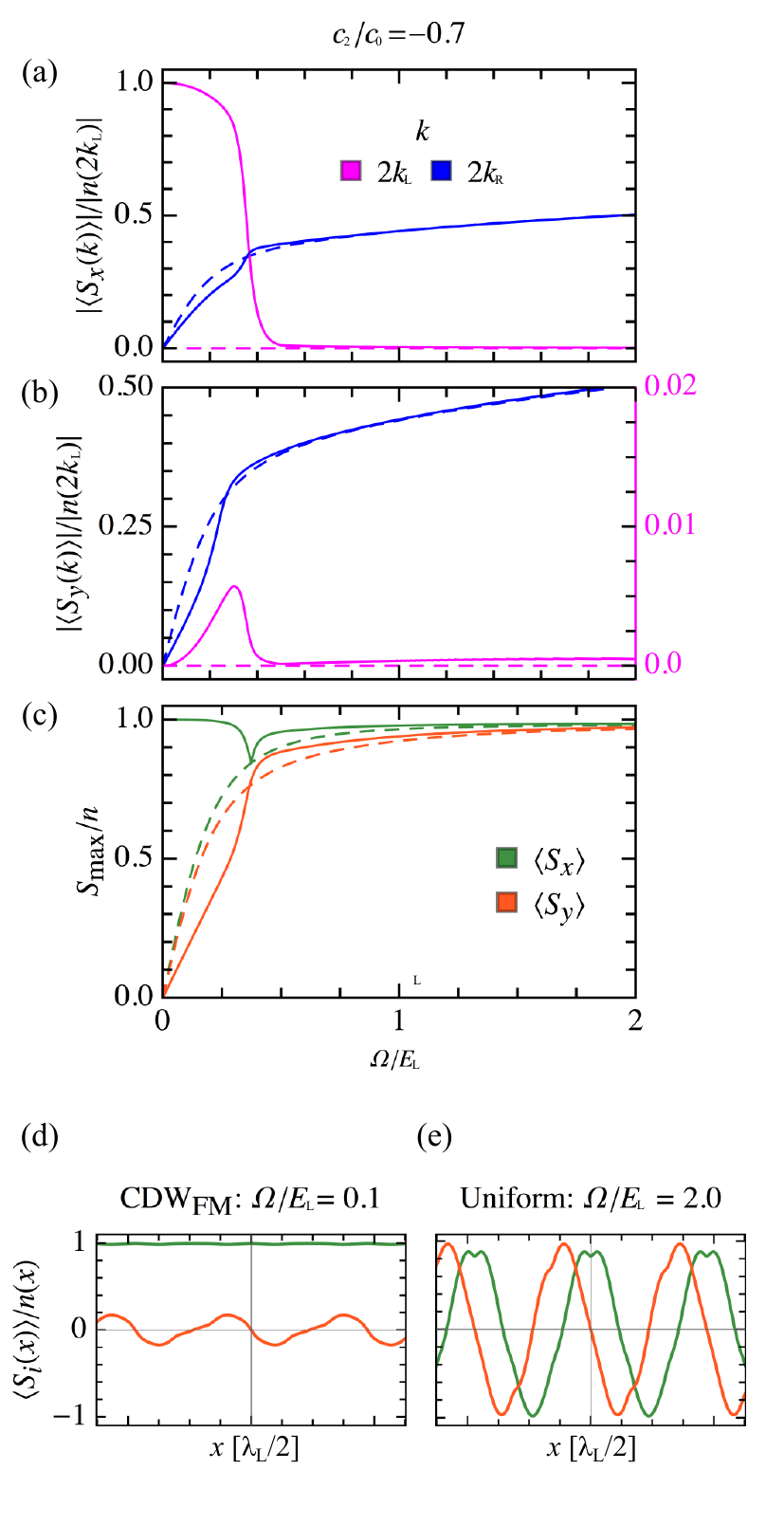}
\caption{(color online)~Spin wave order for $c_2/c_0 = -0.7$, $\varepsilon = 0$. Dashed lines show $c_2, c_0 = 0$ case for reference.~(a)~$\langle S_x(k) \rangle$ is pinned to the density in the CDW$_\mathrm{FM}$ phase, shown by $|\langle S_x(2k_\mathrm{L})\rangle | \neq 0$. After the transition $\langle S_x\rangle $ is modulated primarily at the Raman wavevector $k = 2k_\mathrm{R}$ and $|\langle S_x(2k_\mathrm{L})\rangle |\rightarrow 0$.~(b)~$\langle S_y(k)\rangle $ is nearly unaffected by the optical lattice but follows the Raman beam, shown by $|\langle S_y(2k_\mathrm{L})\rangle | \ll  |\langle S_y(2k_\mathrm{R})\rangle |$.~(c)~Amplitude of spin oscillations with increasing $\Omega$. We see that the ferromagnetic state crosses over to Raman polarized at $\Omega \approx 0.5E_\mathrm{L}$.~(d-e)~Example real-space spin texture.~(d)~CDW$_\mathrm{FM}$ Phase.~(e)~Uniform density phase with a helical spin texture.\label{Fig:SpinF}}
\end{figure}

\section{Results: Attractive Spin-Dependent Interactions}

The schematic phase diagram for $c_2<0$ and $\varepsilon=0$ is shown in Fig.~\ref{Fig:Main}(a). For $\Omega = 0$ the system is an SU(2) FM~\cite{Ho1998,Ohmi1998} with uniform charge density (i.e. the density is only suppressed by the lattice potential); this symmetry is broken by the Raman field. The physics at large $\Omega$ is largely explained by the single-particle Hamiltonian: The data matches up with exact diagonalization for $c_0 = c_2 = 0$ rather well. 

For small $\Omega$, the SU(2) FM phase and the modulating Raman field compete. In this regime, the charge redistributes itself to accommodate the FM phase in the presence of the Raman field. To understand this, consider the Raman field on each lattice site: The angle of the Raman field in the $S_x$-$S_y$ plane is $0$, $2\pi/3$, and $4\pi/3$ before it repeats itself every third site.
Within the GPE, the wave function at each site $l$ is $\psi_{\alpha,l} = \sqrt{n_l} \xi_{\alpha,l}$ with $\xi_{\alpha,l}^\dagger \xi_{\alpha,l} = 1$, and to leading order in $\Omega$ the energy is changed by roughly
\begin{align}
    \Delta E \approx \sum_l n_l~\xi_{\alpha,l}^\dagger\mathbf{F}_{\alpha\beta}\xi_{\beta,l} \cdot \bm \Omega_\mathrm{R}.
    \label{eq:rough_energetics_FM}
\end{align}
For a given FM state, $\Delta E$ is minimized by larger density $n_l$ at sites where $\xi_{\alpha,l}^\dagger\mathbf{F}_{\alpha\beta}\xi_{\beta,l}\cdot \boldsymbol{\Omega}_\mathrm{R}< 0$ and smaller $n_l$ when $\xi_{\alpha,l}^\dagger\mathbf{F}_{\alpha\beta}\xi_{\beta,l} \cdot \boldsymbol{\Omega_\mathrm{R}} > 0$. This reasoning leads to two kinds of charge density waves as depicted in Fig.~\ref{Fig:CDWSketch}. Our simulations suggest that Fig.~\ref{Fig:CDWSketch}(b) is lower energy, shown in Fig.~\ref{Fig:DensityF}(a). We precisely track this CDW$_{\mathrm{FM}}$ regime by looking at Fourier modes of the density $n(k)$ at $k=2k_\mathrm{R}$ as seen in Fig.~\ref{Fig:DensityF}(c), which makes clear that the CDW is an interaction induced effect, increased by a larger FM interaction.

\begin{figure}[t!]
\centering
\includegraphics[width = \columnwidth]{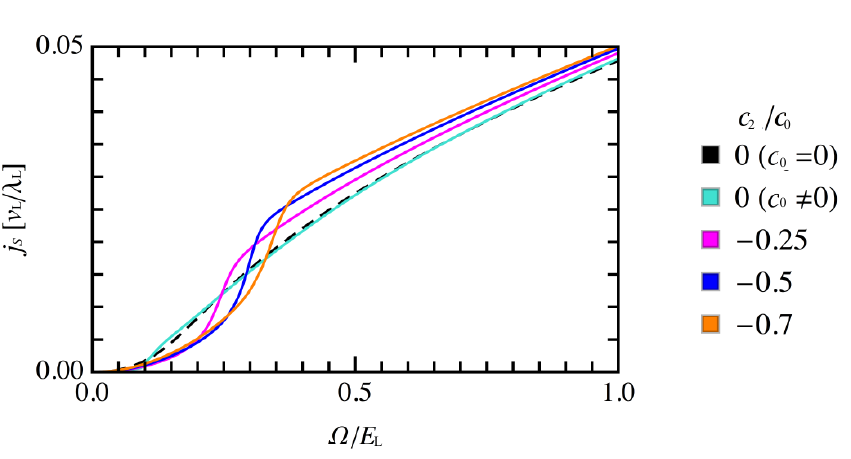}
\caption{(color online)~$j_S(\Omega)$ computed for $c_2/c_0 \leq 0$ and $\varepsilon = 0$. Ferromagnetic spin-dependent interactions first suppress and then slightly enhance the overall spin current compared to the non-interacting case (dashed line) in the regime where the density modulation is highest. For spin-independent interactions only ($c_2 = 0$) the current is hardly changed.\label{Fig:CurrentF}}
\end{figure}

In Fig.~\ref{Fig:SpinF} we describe a system with $0< \Omega/E_{\mathrm{L}} \lesssim 0.5$ that is polarized along $S_x$, connected to two other degenerate states with the transformation
\begin{align}
    \psi(x) \rightarrow e^{-\frac{2\pi i}{3} F_z}\psi(x-\pi/k_{\mathrm{L}}).
    \label{eqn:shift}
\end{align}
We confirmed numerically that the above transformation yields degenerate states with the same energy, and one would expect this from the above reasoning. The precise nature of the FM state with increasing $\Omega$ is captured in Fig.~\ref{Fig:SpinF} where we see that $\langle S_x(k)\rangle $ is at first only modulated at $2k_{\mathrm{L}}$ and $\langle S_y(k)\rangle $ is quite small, but as the Raman field $\Omega$ increases we obtain a small SW in $\langle S_y(x)\rangle$, shown in Fig.~\ref{Fig:SpinF}~(c,d). This minimizes the energy in Eq.~\eqref{eq:rough_energetics_FM}, as depicted in Fig.~\ref{Fig:CDWSketch}.

For strong enough $\Omega$, the easy axis ferromagnetic order is suppressed and the helical order of the single particle picture takes over, as indicated in Fig.~\ref{Fig:Fractions}(c) by the increase in occupation in the $m_F = 0$ state for $\Omega \approx 0.3E_\mathrm{L}$. Finally, a cross-over to a helical spin texture occurs for large $\Omega$, shown in Fig.~\ref{Fig:SpinF}. The preference for $m_F=0$ is seen in the single particle picture: for small $\Omega$ the degenerate $m_F=-1,0,1$ states split so that $m_F=0$ becomes the lowest energy, as discussed in Sec. II. 

The spin-current is initially suppressed by these FM interactions as seen in Fig.~\ref{Fig:CurrentF}. In the synthetic space, a FM state implies little phase change between neighboring sites with $m=1$ or $m=-1$, leading to a suppressed spin current, but a SW is induced as a function of increasing Raman strength. This leads to an increased spin current, and even enhances it past the single particle value where density modulation is highest. Finally, the spin current approaches the non-interacting case.

\begin{figure}[t!]
\centering
\includegraphics[width = \columnwidth]{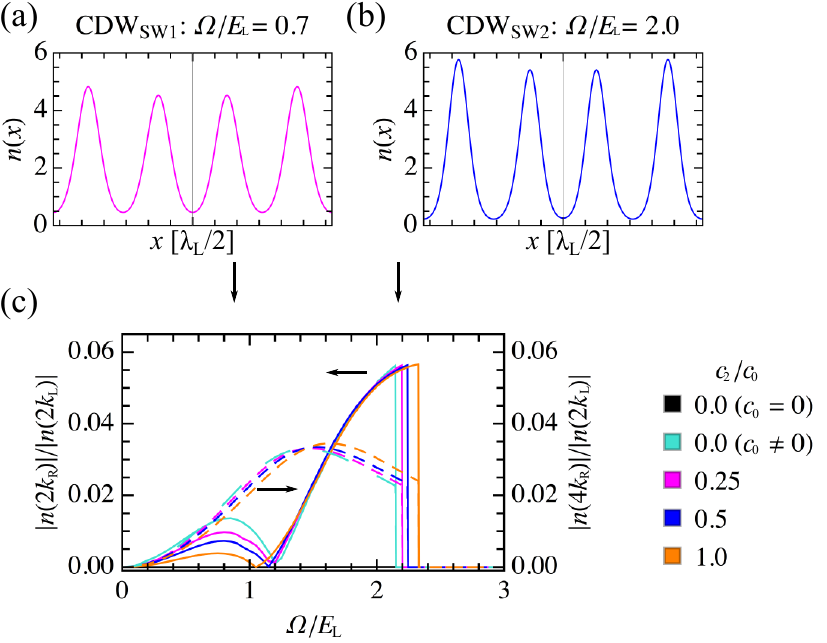}
\caption{(color online)~(a-b)~GPE density computed for $c_2/c_0 = 0.25$ and $\varepsilon = -E_\mathrm{L}$.~(a)~Density in real space in the CDW$_\mathrm{SW1}$ and ~(b)~CDW$_\mathrm{SW2}$ phases.~(c)~$n(k)$ at $k = 2k_\mathrm{R}$ (solid lines), $4k_\mathrm{R}$ (dashed lines). The non-interacting case (black line) has no density modulation other than by the lattice, so it is zero in this case. For $c_0 \neq 0$ the density is modulated at two different wavevectors of the same order of magnitude, varying slightly with varying $c_2$. For $|n(4k_\mathrm{R})| > |n(2k_\mathrm{R})|$ we denote the CDW$_\mathrm{SW1}$ phase, while for $|n(4k_\mathrm{R})| < |n(2k_\mathrm{R})|$ the system is in the CDW$_\mathrm{SW2}$ phase. The cross over from CDW$_\mathrm{SW1} \rightarrow$ CDW$_\mathrm{SW2}$ occurs for $\Omega \approx 1.7E_\mathrm{L}$. The system undergoes a first order transition to uniform density for $\Omega \approx 2.4E_\mathrm{L}$, corresponding to the transition to the single minimum regime.\label{Fig:DensityP}}
\end{figure}

\section{Results: Repulsive Spin-Dependent Interactions}

In the case of polar interactions the system cannot lower its energy through the interplay of enhanced density modulation and spin wave order. Setting $\varepsilon = 0$ gives a uniform density ground state with $n(x)\sim \cos^2(k_\mathrm{L}x)$. An experimentally accessible way to stabilize a CDW phase in a polar BEC is to bias the system with a large negative $\varepsilon$. Setting $\varepsilon = -E_\mathrm{L}$ for $c_2 > 0$ favors occupation of the single-particle minima at maximal spin states. This induces competition between the spin-dependent interaction and the underlying single particle dispersion; in the language of synthetic dimensions, $\varepsilon < 0$ favors edge over bulk states. The phase diagram for $c_2 - \Omega$ and $\varepsilon < 0$ is shown in Fig.~\ref{Fig:Main}(b). There exist two phases with nonzero $n(2k_\mathrm{R})$ and $n(4k_\mathrm{R})$. These phases are denoted CDW$_\mathrm{SW1}$, for $|n(4k_\mathrm{R})| > |n(2k_\mathrm{R})|$, and  CDW$_\mathrm{SW2}$ for $|n(4k_\mathrm{R})| < |n(2k_\mathrm{R})|$, which are analogous to the distinct density modulated phases found without the optical lattice in Ref.~\cite{Yu2016}. The existence of multiple density wave phases allows for the possibility of observing a continuous CDW$_\mathrm{SW1}$ $\rightarrow$ CDW$_\mathrm{SW2}$ cross over with increasing $\Omega$, as shown in Fig.~\ref{Fig:DensityP}(a-c). For $c_2/c_0=0.25, 0.5$ and $1.0$, the cross over occurs at $\Omega \approx 1.7E_\mathrm{L} $. We find a first order phase transition at $\Omega \approx 2.4E_\mathrm{L} $ into the uniform density phase, which occurs when the lowest band becomes extremely flat (not shown). The first order transition is a generic feature of the transition from two minima to one and has also been predicted for the interacting system without the optical lattice~\cite{Yu2016,Martone2015}.

\begin{figure}[t!]
\centering
\includegraphics[width = \columnwidth]{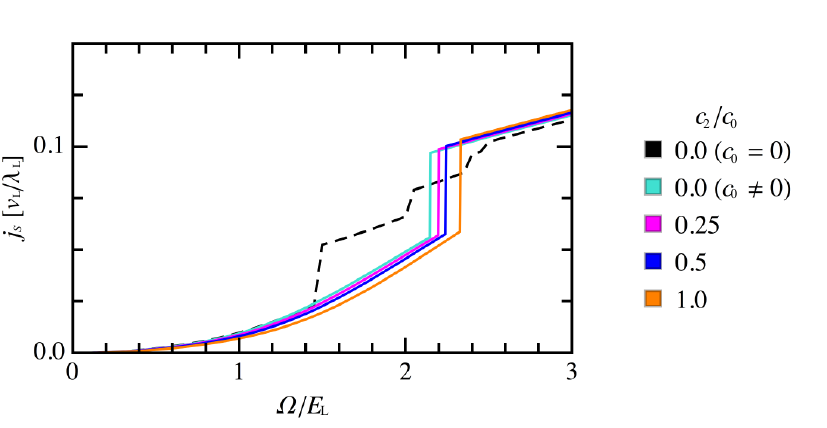}
\caption{(color online)~$j_S(\Omega)$ computed for $c_2/c_0 \geq 0$ and $\varepsilon = -E_\mathrm{L}$. Repulsive spin-independent ($c_0$) interactions suppress current compared to the non-interacting case. The first order transition causes a sharp increase in current, and is weakly dependent on $c_2$. In the non-interacting case (dashed line), the discrete steps in the current are a finite size effect due to the change in curvature of the lowest band. As $\varepsilon$ is tuned, the momentum $k$ where the band minimum occurs decreases in discrete steps from the original value of $k=\pm2k_\mathrm{R}$ until the single minimum regime at $k=0$ is reached. In the infinite system this curve would be smooth.\label{Fig:CurrentP}}
\end{figure}

\begin{figure}[t!]
\centering
\includegraphics[width = 0.85\columnwidth]{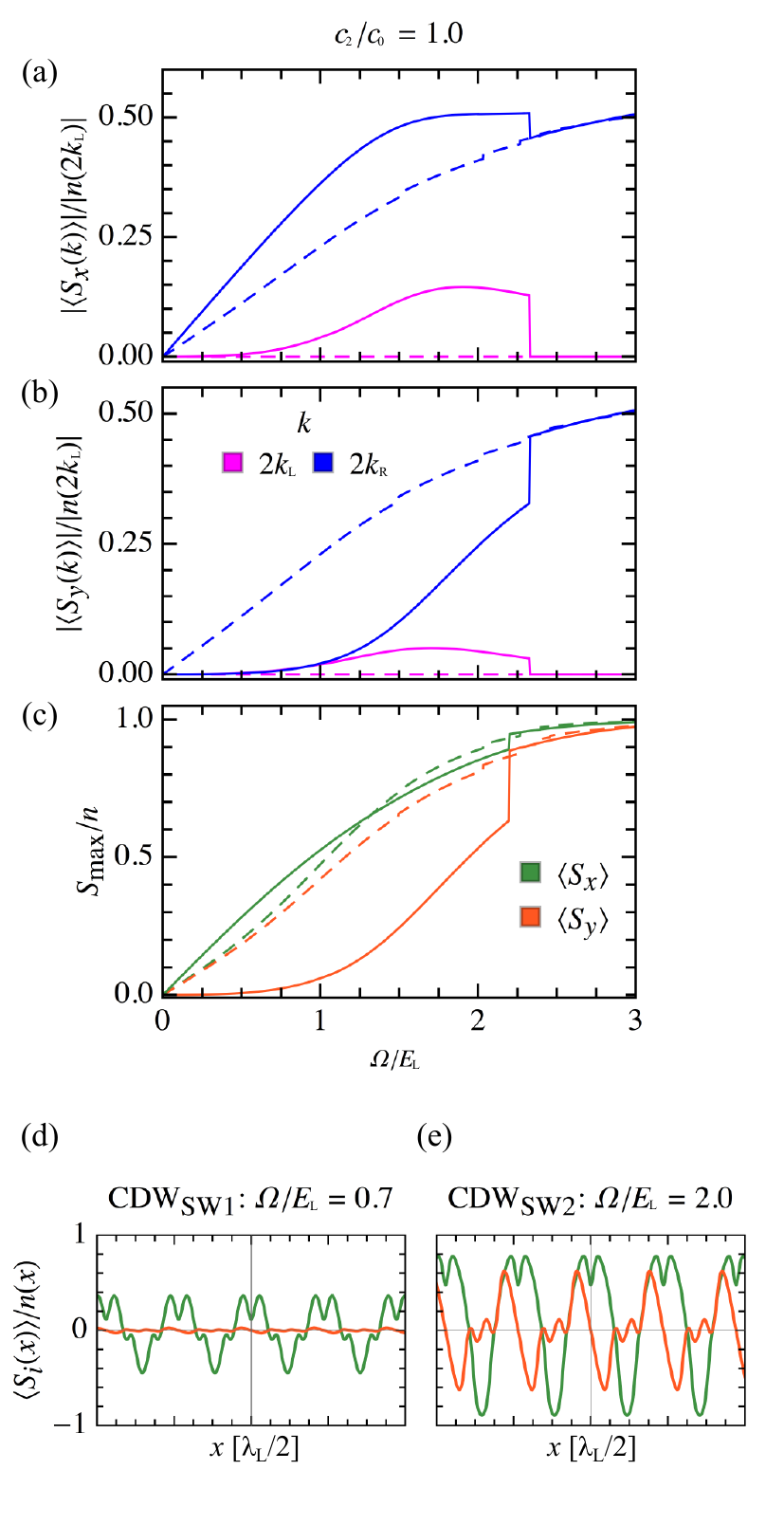}
\caption{(color online)~Spin wave order for $c_2/c_0 = 1.0$, $\varepsilon = -E_\mathrm{L}$. Dashed lines show the $c_2, c_0 = 0$ case.~(a-b)~$\langle S_x(k)\rangle$ and $\langle S_y(k)\rangle $ are modulated at both the Raman and lattice wavevectors for $\Omega \lesssim 2.4E_\mathrm{L}$. After the first order transition the spin comes unpinned from the lattice as evidenced by $|\langle S_{x,y}(k=2k_\mathrm{L})\rangle | = 0$.~(c)~The amplitude of spin oscillations grows with increasing $\Omega$. The non-interacting case initially occupies a single minimum and is polarized in $\langle S_z\rangle $ for $\Omega = 0$ (not shown). For $\Omega \neq 0$ $\langle S_x\rangle$ and $\langle S_y\rangle $ grow continuously, with $\langle S_y\rangle $ suppressed in the interacting case. (d-e) Real-space spin textures for CDW$_\mathrm{SW1}$ and CDW$_\mathrm{SW2}$.\label{Fig:SpinP}}
\end{figure}

Even for the relatively shallow lattice at $V = 5E_\mathrm{L}$ contrast of the CDW phases may be difficult to resolve in experiments. The first order transition can be verified through measurement of the current, shown in Fig.~\ref{Fig:CurrentP}. Spin current is suppressed for $c_2 > 0$ in comparison to the non-interacting case, particularly in the flat-band region around $\Omega \approx 2.0E_\mathrm{L}$. The first order transition leads to a discontinuous jump in spin current. Past this point the spin current approaches the single particle case.

In addition to novel CDW behavior, the system exhibits multiple spin textures. Total spin $\langle\vec{S}^2(x)\rangle$ is minimized for $\Omega = 0$. As $\Omega$ grows the spin begins to polarize in the $S_x-S_y$ plane. Initially $\langle S_y\rangle $ is suppressed and only begins to grow after the CDW$_\mathrm{SW1}$ $\rightarrow$ CDW$_\mathrm{SW2}$ cross over, as shown in Figs.~\ref{Fig:SpinP}(b,c). This spin configuration is also connected to two other degenerate states through the transformation in Eq.~\eqref{eqn:shift}, which we have verified numerically. The lattice plays a much smaller role in the spin textures than in the $c_2 < 0$ case as evidenced by a small but nonzero $|\langle S_{x,y}(2k_\mathrm{L})\rangle |$. At large $\Omega$ the helical spin texture is again entirely determined by the Raman beam, decoupled from the density behavior and the sign of $c_2$. The variety of spin textures is shown in Fig.~\ref{Fig:SpinP}(d,e), where we plot the spin in real space for each of the density wave phases. In the uniform density phase the spin texture is the same as Fig.~\ref{Fig:SpinF}(e). Due to the high degree of degeneracy in the non-interacting case with $\varepsilon = -E_\mathrm{L}$, in Fig.~\ref{Fig:DensityP}-\ref{Fig:CurrentP} we present the non-interacting results for condensation in a single minimum.

\section{Discussion and Conclusion}

We examined the weakly interacting spin-1 Bose gas with SOC in an optical lattice and related it to the synthetic dimensions framework. Specifically, we have presented the phase diagram in the $\Omega-c_2$ plane for both positive and negative values of $c_2$. The system exhibits a rich phase diagram with CDW and SW phases, which depend strongly on $\Omega$, $\varepsilon$, and the sign of $c_2$. In the regime of intermediate lattice depth at $V = 5E_\mathrm{L}$, we find a number of novel phases. For attractive spin-dependent interactions, the system exhibits ferromagnetic behavior and density modulations at the Raman wavevector, leading to altered spin current in the CDW$_\mathrm{FM}$ regime. In the CDW$_\mathrm{FM}$ phase there are small spin modulations that cross over to helical polarization.

BECs with repulsive spin-dependent interactions present novel phases provided that $\varepsilon <0$. In particular,  a cross over from a CDW$_\mathrm{SW1}$ to CDW$_\mathrm{SW2}$ phase occurs with increasing the Raman intensity, and a first order transition to a uniform density state is also seen. This first order transition can be measured through the spin current, which shows a discontinuous jump at the transition. We show that $\varepsilon$ plays a crucial role in the phases that can be realized. Increasing $\Omega$ leads to condensation only at the lattice wavevector, which indicates a uniform density state.

Finally, we studied the interplay of spin and density order parameters by characterizing the spin textures in the $S_x-S_y$ plane. Notably, the interplay of interactions and single particle physics at low $\Omega$ leads to alterations of the spin current and spin texture when compared to the non-interacting case. This is true of both attractive and repulsive $c_2$, however for $c_2 < 0$ this effect is particularly pronounced due to the appearance of finite magnetization. Interactions lead to modified bulk or edge occupation of the system in the synthetic dimension. For $c_2>0$ and large negative $\varepsilon$ the system favors edge occupation of the synthetic lattice, while for $c_2<0$, $\varepsilon = 0$ it is primarily bulk occupation. Future research could investigate the excitation spectrum in the present setup, or more closely examine the role of $\varepsilon$. These results are accessible by current experiments and apply to a variety of atoms such as $^{87}$Rb, $^{7}$Li, $^{41}$K, and $^{23}$Na.

\section*{Acknowledgements}

We would like to thank William Cole and Sankar Das Sarma for useful discussions as well as collaborations on related work. This work is partially supported by the NSF through the PFC at the JQI via the PFC seed grant ``Emergent phenomena in interacting spin-orbit coupled gases''  (HMH, JHP, SSN) for support, JQI-NSF-PFC, LPS-MPO-CMTC, and Microsoft Q (JHP and SSN). HMH acknowledges additional fellowship support from the National Physical Science Consortium and NSA. Additional support provided by the ARO’s Atomtronics MURI, and by the AFOSR's Quantum Matter MURI, NIST, and the NSF through the PFC at the JQI (IBS, JHW). 

\section*{Appendix: Effect of increasing lattice depth}

\begin{figure}[t!]
\centering
\includegraphics[width =0.75\columnwidth]{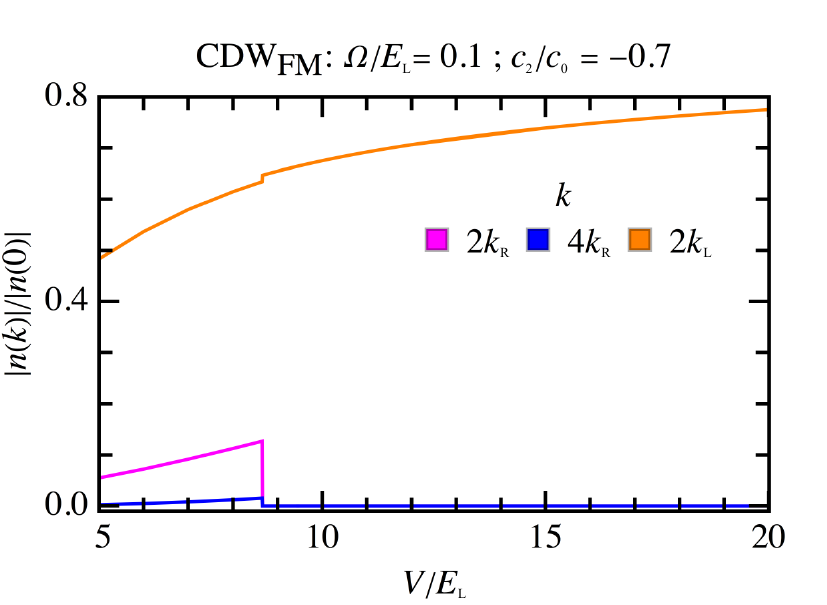}
\caption{(color online)~Dependence of the CDW$_\mathrm{FM}$ phase on lattice depth. As $V_\mathrm{L}$ increases, the CDW increases in amplitude and then undergoes a first order transition to a uniform density phase. Condensation moves toward the Brillouin zone edge, as shown by increasing $|n(2k_\mathrm{L})|$ (orange line) even after the transition. Note that here $|n(k)|$ is normalized by $|n(k = 0)|$.\label{Fig:VLFM}}
\end{figure}

In this Appendix, we analyze the role of increasing lattice depth $V_\mathrm{L}$ on the CDW phases presented in the main body of the paper. The mean-field description of the BEC breaks down as the lattice depth increases and Mott physics becomes more important. In our results, we find that for both polar and ferromagnetic spin-dependent interactions the CDW order is suppressed for $V_\mathrm{L}\gtrsim 10E_\mathrm{L}$. This indicates that increasing the lattice depth by as little as a factor of two reaches the boundary of applicability of the mean-field description in 1D.

\begin{figure}[b!]
\centering
\includegraphics[width =0.75\columnwidth]{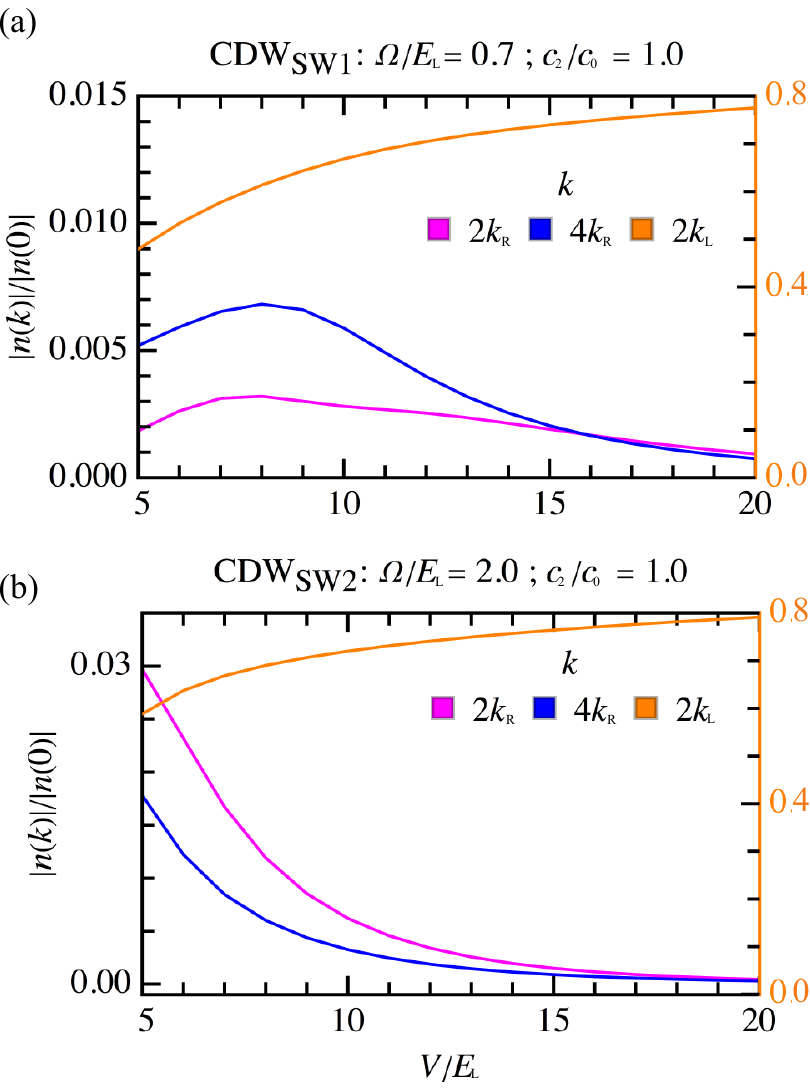}
\caption{(color online)~(a-b)~Dependence of the CDW$_\mathrm{SW1}$ and CDW$_\mathrm{SW2}$ phases on lattice depth. In both cases, condensation moves toward the Brillouin zone edge, as shown by increasing $|n(2k_\mathrm{L})|$ (orange line, right axis), which is much larger in magnitude than the other order parameters.~(a)~CDW$_\mathrm{SW1}$. As $V_\mathrm{L}$ increases, the CDW increases slightly before decreasing.~(b)~CDW$_\mathrm{SW2}$. As $V_\mathrm{L}$ increases, the CDW decreases. Notably, $|n(2k_\mathrm{L})|$ is the same order of magnitude as the CDW$_\mathrm{FM}$ case, while $|n(2k_\mathrm{R})|$ and $|n(4k_\mathrm{R})|$ are much smaller. Note that in both (a) and (b) $|n(k)|$ is normalized by $|n(k = 0)|$.\label{Fig:VLSW}}
\end{figure}

In Figure \ref{Fig:VLFM} we analyze the effect of an increasing lattice depth on the CDW$_\mathrm{FM}$ phase. As $V_\mathrm{L}$ increases, the CDW amplitude  \emph{grows} until there is a first-order transition to a uniform density phase with modulation only at $k = 2k_\mathrm{L}$. The transition occurs for $V_c \approx 8.6E_\mathrm{L}$, therefore it is important that the system is in a relatively shallow lattice regime to observe the CDW$_\mathrm{FM}$ phase. For $V_\mathrm{L}>V_c$, condensation only occurs at $k=0$ and the Brillouin zone edge ($k=k_L$) with increasing lattice depth, as shown by an increase in $|n(2k_\mathrm{L})|$ with increasing $V_\mathrm{L}$. The deep lattice suppresses interaction-induced effects at the mean-field level (for these values of $c_0$ and $c_2$) including the CDW and FM polarization, and the spin texture is Raman-polarized as the lattice depth increases.

The effect of increasing lattice depth on both CDW$_\mathrm{SW1}$ and CDW$_\mathrm{SW2}$ is shown in Figure \ref{Fig:VLSW}. Unlike the CDW$_\mathrm{FM}$ case, the amplitude of density modulations decreases gradually with increasing $V_\mathrm{L}$ and we do not find a sharp transition. For CDW$_\mathrm{SW1}$ the behavior of $n(k)$ at $k = 2k_\mathrm{R}$ and $4k_\mathrm{R}$ is slightly non-monotonic, showing a small increase initially with increasing $V_\mathrm{L}$. In the case of CDW$_\mathrm{SW2}$, $n(k)$ decreases across the entire range of $V_\mathrm{L}$. Condensation at the lattice wavevector is almost  the same as for the CDW$_\mathrm{FM}$ case, with $n(2k_\mathrm{L})$ having a similar magnitude. This shows that for increasing lattice depth the condensation wavevector moves to the edge of the Brillouin zone independent of the type of interactions present. 

\newpage

\bibliography{main}

\end{document}